\documentclass[a4,9pt]{article}
\usepackage{psfig}
\def\micron{$\mu$m}
\def\msunyr{\ifmmode M_{\odot} {\rm yr}^{-1} \else M$_{\odot}$ yr$^{-1}$\fi}
\newcommand{\la}{\raisebox{-0.5ex}{$\,\stackrel{<}{\scriptstyle\sim}\,$}}
\newcommand{\ga}{\raisebox{-0.5ex}{$\,\stackrel{>}{\scriptstyle\sim}\,$}}

\begin{document}
\centerline{{\bf ESO Messenger article, september 2006 issue (No. 125, p. 20)}}
\centerline{Text and full resolution version at: http://www.eso.org/gen-fac/pubs/messenger/}
\vspace{1cm}

\section*{Searching for the first galaxies through gravitational
lenses}
{\em Daniel Schaerer$^{1,2}$, 
Roser Pell\'o$^2$,
Johan Richard$^{2}$,
Eiichi Egami$^{3}$,
Angela Hempel$^{1}$,
Jean-Fran\c{c}ois Le Borgne$^2$,
Jean-Paul Kneib$^{4,5}$,
Michael Wise$^6$,
Fr\'ed\'eric Boone$^7$,
Fran\c{c}oise Combes$^7$
}

$^1$ Geneva Observatory, 51 Ch. des Maillettes, CH--1290 Sauverny, Switzerland,
$^2$ Observatoire Midi-Pyr\'en\'ees, Laboratoire d'Astrophysique, UMR 5572, 
	14 Avenue E. Belin, F-31400 Toulouse, France,
$^3$ Steward Observatory, University of Arizona, 933 North Cherry Avenue, Tucson, AZ
                     85721, USA,
$^4$ OAMP, Laboratoire d'Astrophysique de Marseille, UMR 6110 traverse du Siphon, 13012 Marseille, France,
$^5$ Caltech Astronomy, MC105-24, Pasadena, CA 91125, USA,
$^6$ Astronomical Institute Anton Pannekoek, Kruislaan 403, NL-1098 SJ Amsterdam, The Netherlands,
$^7$ Observatoire de Paris, LERMA, 61 Av. de l'Observatoire, 75014 Paris, France

{\bf 
Observing the first galaxies formed during the reionisation epoch,
i.e. approximately within the first billion years after the Big Bang,
remains one of the challenges of contemporary astrophysics.
Several efforts are being undertaken to search for such remote objects.
Combining the near-IR imaging power of the VLT and the natural
effect of strong gravitational lensing our pilot program has allowed
us to identify several galaxy candidates at redshift $6 \la z \la 10$.
The properties of these objects and the resulting constraints
on the star formation rate density at high redshift are discussed.
Finally we present the status of follow-up observations 
(ISAAC spectroscopy, HST and Spitzer imaging) and discuss 
future developments.
}

Like the explorers of seas and continents in the past centuries,
astronomers keep pushing the observational frontiers of the universe
with their telescopes thereby tracing back the history of stars and
galaxies since their birth. Just a few years ago, in 2002, the most
distant galaxy known was a faint unconspicuous object called HCM 6A at
redshift $z=6.56$ discovered thanks to the natural effect of
gravitation lensing provided by a foreground cluster of galaxies,
which magnifies the light of this distant background object (Hu et
al. 2002).  Few quasars at similar distance had also been discovered,
pushing our ``observational horizon'' already back
in time to a little less than 1 billion years after the Big Bang,
close to the end of cosmic reionisation.  However, evidence from
cosmic microwave background polarisation measurements and other
arguments indicate that star formation in galaxies must have occurred
even earlier.

Since the late nineties, the so-called Lyman break or ``dropout''
technique had established itself as a simple and successful method to
identify distant galaxies through the use of broad band photometry.
Furthermore sensitive infrared instruments
were now available on the large ground-based telescopes, able in
principle to detect proto-galaxies out to even higher redshifts.
Given these
advances, we started few years ago a pilot-project with the aim of
finding star-forming galaxies at redshifts beyond 6-6.5 using lensing
clusters as gravitational telescopes, a well-established technique
nowadays.  This project is mainly based on ISAAC and FORS2 data plus
additional observations obtained at CFHT and HST.

\subsection*{Observations}
Focussing on two well known gravitational lensing clusters, Abell 1835 and 
AC114, we obtained deep near-IR images in the $SZ$, $J$, $H$, and $Ks$ bands
with ISAAC and an additional $z$-band image with FORS2.
These deep images, reaching e.g. a $1 \sigma$ depth of 26.1 in $H_{AB}$, 
were then used to search for objects which are detected at least in two near-IR
bands, which show a blue near-IR colour, and which are undetected (i.e. ``dropped 
out'') in all optical bands. 
These criteria are optimised to select high redshift ($z > 6$) objects
with intrinsically blue UV-restframe spectra, i.e. very distant starburst galaxies,
and to avoid contamination from intrinsically faint and red cool stars.
Different combinations of colour-colour plots allow a crude classification
into several redshift bins. An example of such a diagram, showing the expected
location of $z \sim$ 8--11 galaxies and of candidates found behind Abell 1835
is shown in Fig. 1. A complete report is given in Richard et al. (2006).

\subsection*{High-z galaxy candidates and the cosmic star formation
density during reionisation}
In spring 2003 the analysis of the candidates behind Abell 1835
yielded an intriguing, strongly lensed object whose spectral energy
distribution was compatible with that of a galaxy at $z \sim$ 9--11.
During our first spectroscopic follow-up run with ISAAC in summer 2003
we were able to secure $J$-band long-slit spectroscopy centered at
this position under excellent seeing conditions.  Interestingly,
careful data reduction revealed at this location the presence of a
single faint emission line detected at $\sim 4-8 \sigma$ (depending on the
integration aperture and stacking procedure), which if 
interpreted as Ly$\alpha$ would indicate $z=10.0$! The report of these
findings has been published in Pell\'o et al. (2004a).

{\em What is the status of this fairly unique candidate?}  Weatherley
et al. (2004) have questioned the reality of the emission line.
However, their negative result could be due to the combination of two
factors: an error in our absolute wavelength calibration discovered
later, and their spectroscopic data reduction technique, where the
information is only preserved at the original (wrong) wavelength
position and smeared elsewhere (see Pell\'o et al. 2004b).  On the
imaging side, deep $V$-band observations with FORS2 (Lehnert et
al. 2005) have confirmed its optical non-detection.  Our ISAAC
$H$-band images have been reanalysed by several groups using different
methods (Bremer et al. 2005, Smith et al. 2006) yielding measurements
compatible with ours typically within 1 $\sigma$.  Surprisingly
however, this object remained undetected in a deeper NIRI/GEMINI
$H$-band image taken approximately 15 months after our ISAAC image
(Bremer et al. 2005).  In spring 2004 we obtained two $SZ$ ($\sim$
1.06 \micron) images with ISAAC, where this object is again detected
(see Pell\'o et al. 2005).  Taken together these spectroscopic and
photometric detections, albeit individually of relatively low
significance, indicate that this source is most likely not a spurious,
but an intrinsically variable object, as discussed in Richard et
al. (2006). Therefore, its nature and precise redshift remain
puzzling, and we exclude this object from our list of high-$z$
candidates, which we will discuss now.

Applying the above selection criteria to the observations of the two
lensing clusters has yielded 13 candidates whose
spectral energy distribution (SED) is compatible with that of star
forming galaxies at $z \ga 6$ (see Richard et al. 2006).  Images and
the SEDs of some of them are shown in Fig. 2 for illustration. The
typical lensing magnification reaches from 1.5 to 10, with an average
of $\sim 6$, i.e. nearly 2 magnitudes. Their star formation rate, as
estimated from the UV restframe luminosity, is typically between $\sim 4$ and
20 \msunyr\ after correcting for lensing.

We have used this data to attempt to constrain for the first time the density
of star-forming galaxies present at $6 \la z \la 10$ using lensing clusters.
After taking into account the detailed lensing geometry, sample
incompleteness, and correcting for false-positive detections we have
constructed a luminosity function (LF) of these candidates assuming a
fixed slope taken from observations at $z \sim 3$.
Within the errors the resulting LF is compatible with that of $z \sim
3$ Lyman break galaxies.  At low
luminosities it is also compatible with the LF derived by Bouwens
and collaborators
for their sample of $z\sim 6$ candidates in the Hubble
Ultra Deep Field (UDF) and related fields. However, the turnover
observed by these authors towards the bright end relative to the
$z\sim 3$ LF is not observed in our sample. 
Finally, from the LF we determine the UV star formation rate (SFR) density
at $z \sim$ 6--10, shown in Fig. 3.
Our values indicate a similar SFR density as between $z \sim$ 3 to 6, 
in contrast to the drop found from the deep NICMOS fields.
Further observations are required to fully understand these differences.
Taken at face value, this relatively high SFR density is in good
agreement e.g. with the recent hydrodynamical models of Nagamine et al.\ (2005),
with the reionisation models of Choudhury \& Ferrara (2005),
and also with the SFR density inferred from the past star formation
history of observed $z \sim 6$ galaxies (e.g. Eyles et al. 2006).

\subsection*{Follow-up observations}
Compared to the dataset just described several additional observations
could be secured on these clusters in the meantime.  For example, deep
$z$-band images of both clusters were obtained with the ACS camera
onboard HST.  These observations confirm that the vast majority
(all except one of the above 13)
of our high-$z$ candidates are optical dropouts as expected,
remaining undetected down to a $1 \sigma$ limiting magnitude of
28.--28.3 mag$_{AB}$ (Hempel et al. 2006).  In collaboration with
Eiichi Egami we have also access to IRAC/Spitzer GTO images at 3.6,
4.5, 5.8 and 8.0 \micron\ of large sample of lensing clusters
including Abell 1835 and AC114.  
Again, none of our high redshift galaxy candidates are detected.
This is easily understood, since extrapolation of their intrinsically blue 
SED to IRAC wavelengths shows that their expected fluxes fall below
the Spitzer sensitivity.
It implies that these objects do not host ``old'' stellar populations
with strong Balmer breaks, and that they are not affected by significant
extinction.
A more detailed account of these observations will be 
presented elsewhere.
For comparison a brighter lensed galaxy at $z \sim 7$ 
has been detected earlier by Spitzer (Egami et al. 2005).

In parallel several attempts were made to detect emission lines from
selected candidates using near-IR long-slit spectroscopy with ISAAC on
the VLT. See Pell\'o et al. (2005) for a preliminary report.
Presently such observations are tedious, fairly time-consuming, and
require excellent seeing conditions. Indeed, given the faintness of
the expected lines, the strong IR background, and the need for
the highest spectral resolution possible to minimise the impact of the
numerous sky lines, a ``scan'' of the entire $J$-band at $R \sim 3000$ 
for example requires 5 settings and a total of $\sim$ 10 ksec
to detect an unresolved line of $(6-8) \times 10^{-18}$ erg s$^{-1}$ cm$^{-2}$ flux
at 5 $\sigma$ with ISAAC.

One or more emission lines could be detected in few objects,
as shown in Fig. 4. For example, one of our secondary targets turned out to be
a very faint $z=1.68$ emission line galaxy.
Other lines are clearly identified as the [O~{\sc ii}] $\lambda$3727,3729 
doublet from an intermediate redshift galaxy. Finally several objects
show a single emission line, which  -- if identified as Ly$\alpha$ -- yields a redshift
compatible with the (high) estimated photometric redshift. 
However, in none of these cases a clear asymmetry of the line, as typical
for Ly$\alpha$ from lower $z$ starbursts, was found. Searches for additional lines
in these objects (e.g.\ C~{\sc iv} $\lambda$1550 or He~{\sc ii} $\lambda$1640
if at high $z$, or for [O~{\sc iii}] $\lambda$5007 or H$\alpha$ if at low $z$)
have so far been negative.
Therefore the redshift of these objects is currently difficult to establish, but 
high-$z$ cannot be excluded on the present grounds.
Forthcoming, more efficient near-IR spectrographs should allow a significant
breakthrough in this field.

\subsection*{Spin-off projects on EROs and dusty intermediate-$z$ galaxies}
Our search for optical dropout galaxies behind lensing clusters yields
also other interesting objects, such as extremely red objects (EROs;
see Richard et al. 2006).  In contrast to the high-$z$ candidates
most of them are detected by IRAC/Spitzer.  
These objects, most likely all at intermediate redshift ($z \sim$ 1--3),
turn out to have similar properties as e.g.\ faint IRAC selected EROs
in the Hubble UDF, 
other related objects
such as the putative post-starburst $z \sim$ 6.5 galaxy of Mobasher et al. (2005),
and some sub-mm galaxies (see Hempel et al. 2006)
Spectroscopic observations with FORS2 and X-ray Chandra observations are also
being secured to clarify the redshift and nature of these interesting optical
dropout objects.

\subsection*{Future}
With our pilot program it has been possible to find several very high
redshift candidate galaxies by combining the power of strong
gravitational lensing with the large collecting area of the VLT.
However, differences with other studies based on deep blank fields are
found, and already differences between our two clusters indicate that
these could at least partly be due to field-to-field variance.  Given
the relatively low S/N ratio of the high-$z$ candidates and the large
correction factors applied to this sample, it is of great interest to
increase the number of lensing clusters observed with this technique.

Furthermore, rapidly upcoming new spectrographs such as the second generation
near-IR VLT instruments XShooter and KMOS, the EMIR 
spectrograph on the Spanish GRANTECAN telescope and others will provide a huge efficiency 
gain for spectroscopic follow-up of faint candidate sources, thanks 
to their increased spectral coverage and multi-object capabilities.
Observations at longer wavelengths, e.g. with HERSCHEL,
APEX and later ALMA, are also planned to search for possible dust emission
in such high-$z$ galaxies and to characterise more completely other populations
of faint optical dropout galaxies. 
Finally the JWST and ELTs will obviously be powerful machines to study
the first galaxies.
Large territories remain unexplored in the early universe!

{\small
\subsubsection*{References}




\hspace*{0.5cm} Bremer, M. et al. 2004, ApJ 615, L1
 

Choudhury, T.R., Ferrara, A. 2005, MNRAS 361, 577

Egami, E., et al. 2005, ApJ 618, L5

Eyles, L., et al. 2006, MNRAS submitted (astro-ph/0607306)


Hempel, A. et al. 2006, A\&A submitted


Hu, E., et al. 2002, ApJ 568, L75

Lehnert, M., et al. 2005, ApJ 624, 80

Mobasher, B., et al. 2005, ApJ 635, 832

Nagamine, K., et al. 2005, New Astronomy Reviews 50, 29

Pell\'o, R., et al. 2004a, A\&A 416, L35

Pell\'o, R., et al. 2004b, astro-ph/0407194 

Pell\'o, R., et al. 2005, IAU Symp. 225, 373


Richard, J., et al. 2006, A\&A in press (astro-ph/0606134)



Smith, G.P, et al. 2006, ApJ 636, 575
 

Weatherley, S.J., et al. 2004, A\&A 428, L29

}

\begin{figure*}[ht]
\centerline{\mbox{\psfig{figure=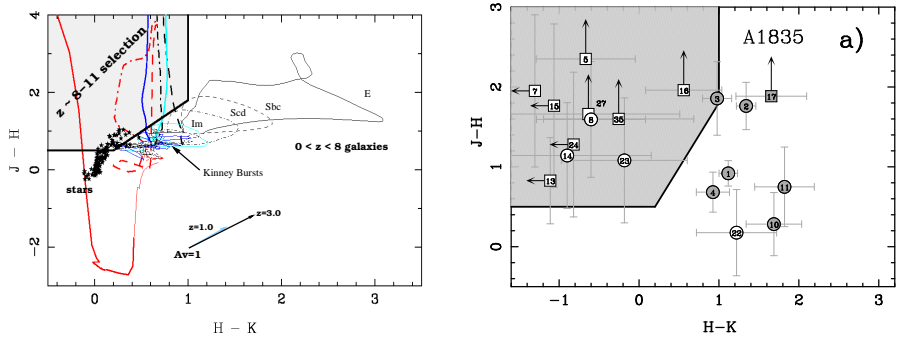,width=5.5cm}\hspace{0.5cm}
\psfig{figure=schaerer_a1835_JHK.ps,height=4.4cm,angle=270}}}
\caption{Colour-colour diagrams (in the Vega system) showing 
{\bf (Left:)} the location for different objects over the 
interval $z \sim$ 0 to 11 and our selection region for galaxies in the
$z \sim$ 8-11 domain and
{\bf (Right:)} the location of the individual optical dropouts detected in Abell 1835.
Circles and squares correspond to high-$z$ candidates detected in 
three and two filters respectively.
Optical dropouts fulfilling the ERO definition are shown in
grey.}
\end{figure*}

\begin{figure}[h]
\psfig{figure=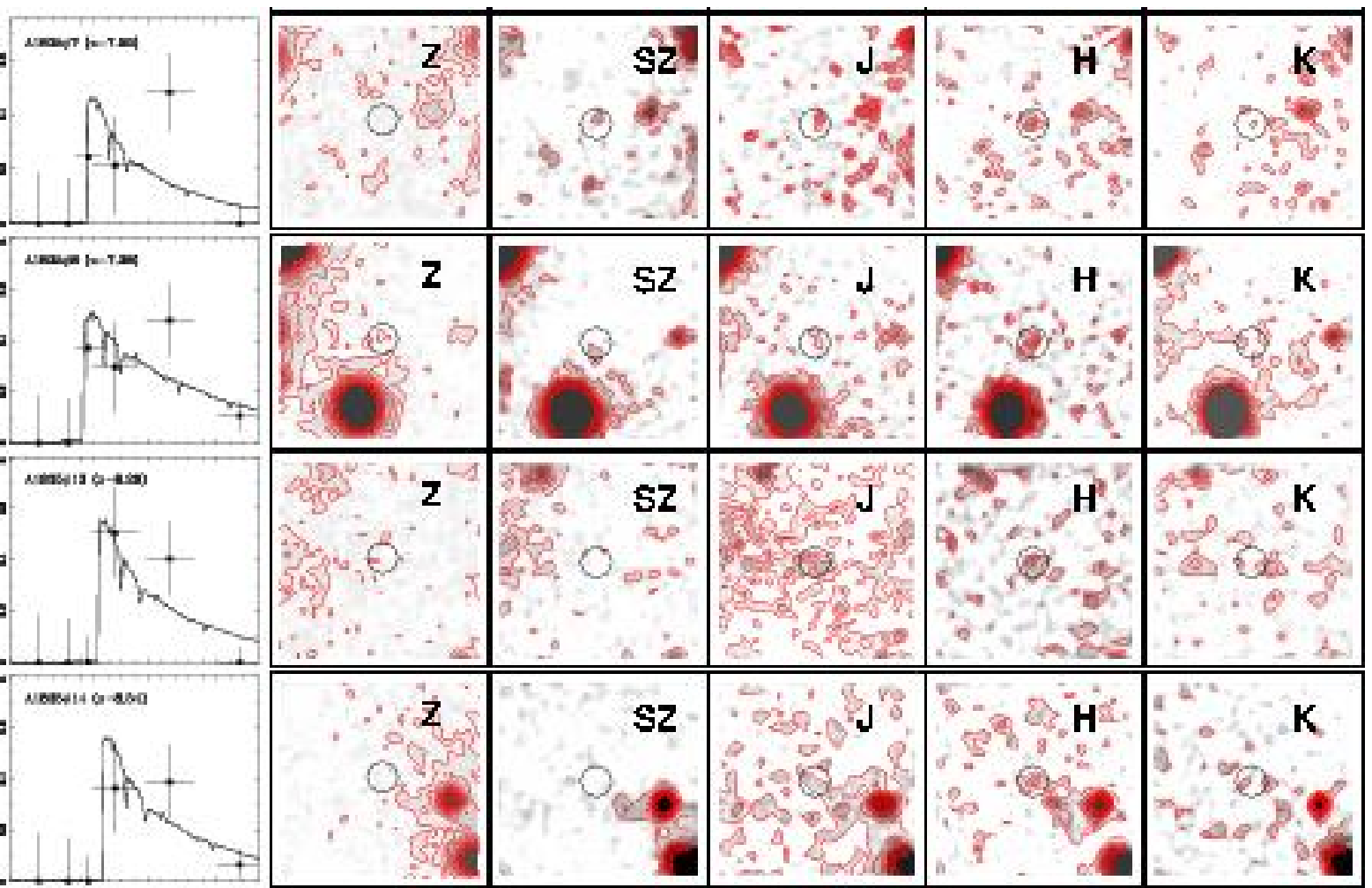,width=12cm}
\psfig{figure=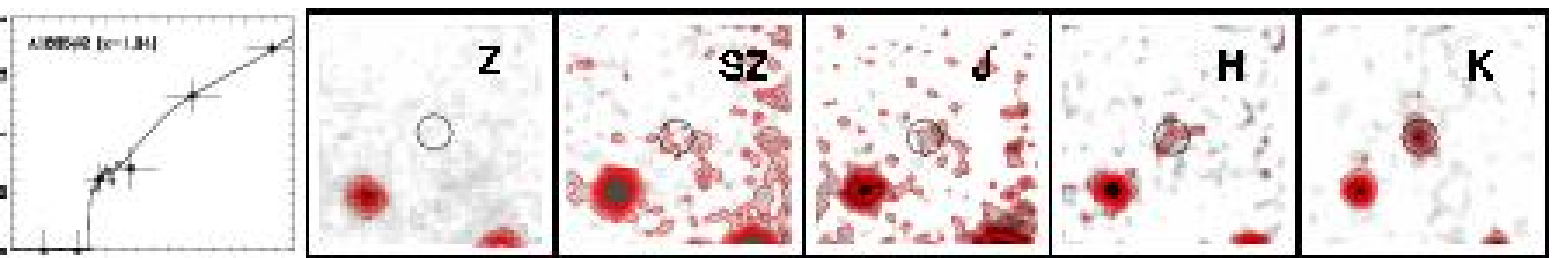,width=12cm}
\caption{Close-up of four high-$z$ candidates in Abell 1835, showing 
the objects and their surrounding 10 $\times$ 10 arcsecs field in
optical ($z$) and near-IR bands ($SZ$, $J$, $H$, and $Ks$) as well as
their SED. For comparison the image and SED of an intermediate
redshift ERO is also shown at the bottom.}
\end{figure}

\newpage

\begin{figure}[htb]
\psfig{figure=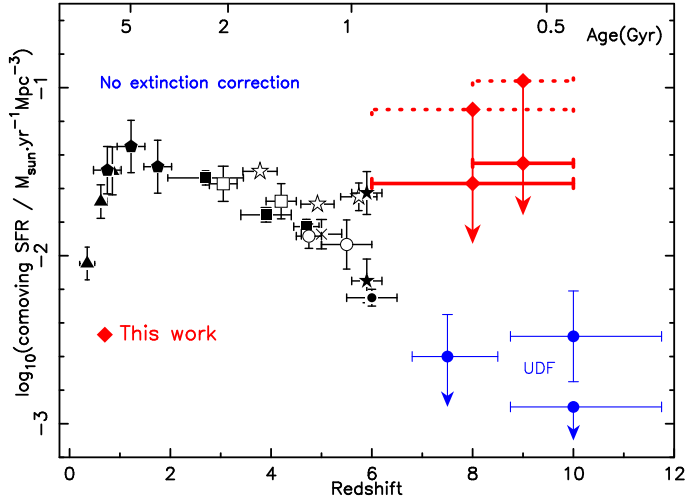,width=9cm,angle=270}
\caption{Evolution of the comoving Star Formation Rate (SFR) density as a function
of redshift including a compilation of results at $z \protect\la 6$, 
our estimates obtained from both clusters for the redshift 
ranges [$6-10$] and [$8-10$] and the values derived by Bouwens and collaborators
from the Hubble Ultra-Deep Field (labeled ``UDF''; Bouwens et al. 2004, ApJ,
616, L79 and 2005, ApJ, 624, L5).
Red solid lines: SFR density obtained from
integrating the LF of our first category candidates down to $L_{1500}=0.3\ L^{*}_{z=3}$;
red dotted lines: same as red solid lines but including also second category candidates
with a detection threshold of $<2.5 \sigma$ in $H$.}
\end{figure}

\begin{figure}[htb]
\psfig{figure=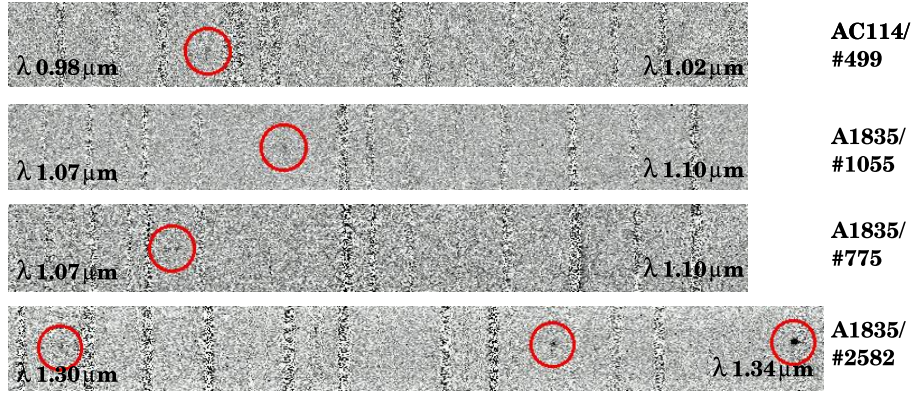,width=12cm}
\caption{Sky-subtracted 2D ISAAC spectra showing examples of objects with emission
line detections marked by red circles in the $J$ band. From top to bottom:
A $z=7.17$ candidate in AC114, a $z=7.89$ candidate in Abell 1835, an intermediate-$z$
galaxy identified by its [O~{\sc ii}] $\lambda$3727,3729 
doublet, and the $z=1.68$ emission line galaxy discovered by 
Richard et al. (2003, A\&A 412, L57). The black vertical lines correspond to sky lines.} 
\end{figure}

\end{document}